\documentclass{Interspeech2024}
\usepackage{multirow}
\usepackage{svg}	
\usepackage{color, colortbl}
\usepackage{hyperref}
 
\definecolor{Gray}{gray}{0.9}

\interspeechcameraready

\title{Study on Inter and Intra Speaker Variability in Speaker Recognition}

\name[affiliation={1}]{Anton}{Okhotnikov}
\name[affiliation={1}]{Nikita}{Torgashov}
\name[affiliation={1}]{Ivan}{Yakovlev}
\name[affiliation={1}]{Pavel}{Malov}
\name[affiliation={1}]{Rostislav}{Makarov}

\address{ID R\&D Inc., New York, USA}
\email{\tt\{ohotnikov,torgashov,yakovlev,pavel.malov,makarov\}@idrnd.net}

\keywords{speaker recognition, session variability, speaker verification}

\begin{document}

\maketitle

\begin{abstract}

Optimization of a trade-off between the number of speakers and their temporal variability (or session diversity) is crucial for the development of a speaker recognition system together with making the data collection process feasible from a time perspective. In this article, we provide the analysis of dependency between inter and intra speaker variability in training data for the modern neural network-based speaker recognition system using the VoxTube dataset for text-independent speaker recognition task. Besides, an auxiliary contribution of this work is a release of upload date metadata per utterance in a VoxTube dataset. We want this article to contribute to guidelines and best practices for collecting and filtering data from media hosting platforms to facilitate the efforts of researchers in developing speaker recognition systems.

\end{abstract}

\section{Introduction}
Nowadays, one of the most popular data collection streams for training machine learning models is the utilization of audio and video hosting platforms \cite{lin2023voxblink, Chung18b, zhu2021webface260m}. Such collection methods show themselves to be easier and cheaper to collect compared to manual data collection, at the same time providing opportunities for process automation and high scaling, hence delivering huge training datasets.

Incorporating both inter and intra speaker variability into training data is essential for the performance of speaker recognition systems, considering both extrinsic and intrinsic factors \cite{HANSEN201894}. High variability of recording sessions (intra speaker) allows for compensation for such extrinsic effects as recording equipment or background noise. Moreover, having such training data is also beneficial for modeling and learning the following intrinsic speaker voice-changing factors: emotional state, health conditions, aging, and speaking style \cite{HANSEN201894}. As a result, high intra speaker variability provides thorough speaker coverage and allows robust speaker embedding modeling, and in the face recognition domain too \cite{facerec_intraspeaker}.

At the same time, a high amount of identities (inter speaker variability) in the training dataset is essential for the performance of any speaker recognition system. It highly improves the model generalization and discrimination capabilities which is confirmed by the results of recent challenges \cite{voxsrc2023_dku_voxblink, voxsrc2023_idrnd_voxtube}. Moreover, there are investigations made in the face recognition field consolidating this statement \cite{zhu2021webface260m}.

The necessity of a study focused on the speaker-session trade-off in speaker recognition arises from the critical need to understand how it impacts the accuracy and robustness of modern speaker recognition systems based on neural networks trained with margin loss. This understanding is particularly crucial for researchers and practitioners collecting and labeling datasets for training and fine-tuning purposes, as it would help to optimize the collection time and data quality. In the end, efficiently collected data is key to robust and resilient across diverse real-world scenarios models and further field progress. Hence, we are interested in answering the following questions in the context of training dataset collection for text-independent (TI) speaker recognition task:
\begin{itemize}
    \item Is high inter speaker variability a key performance factor of a training dataset?
    \item What data setup is more efficient, a high number of speakers with low time variability, or high time variability with less amount of speakers?
    \item How big the intra speaker variability should be (e.g. in years) and what is the optimal amount of recording sessions?
    \item How many speech would be enough to model a speaker and saturate architecture performance?
\end{itemize}

For this work, we justify the selection of a VoxTube dataset \cite{yakovlev23_interspeech} by its large amount of identities, recording sessions per speaker, and their time variability making it a good fit for a problem statement.

The work is done under the assumption that each speaker session is recorded independently according to the video upload date. Due to the incredible difficulty of collecting such data with the years of diversity, we strongly believe that the main conclusions of this research would beneficially scale down to weeks or days too. The limitation of this work is that there is no possibility to control the extrinsic intra speaker factors such as recording equipment or background noise, since no labels are presented in the dataset. As a result, these uncontrolled latent factors may affect the final results.

\begin{figure*}[t]
  \centering
  \includegraphics[width=\linewidth]{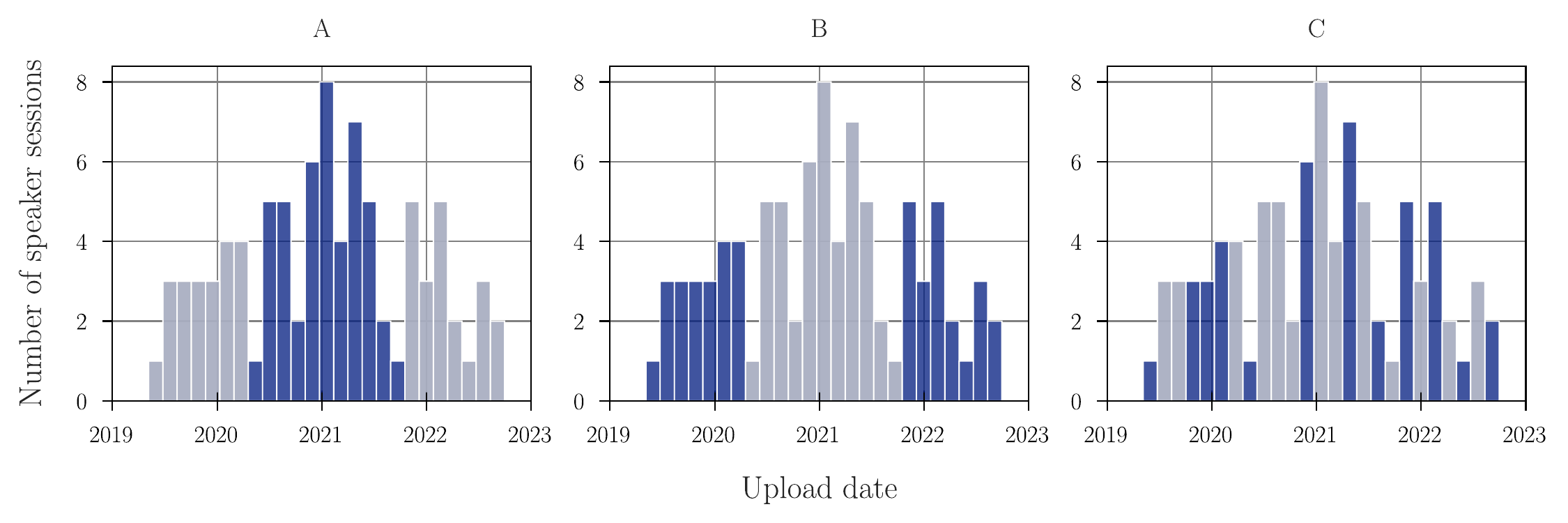}
  \caption{Examples of sessions sampling for experiments \ref{subsection:importance}-\ref{subsection:exps_random_and_adaptive_drop}: A - center sampling, B - tails sampling, C - random sampling. Sessions colored in blue were used in experiments, while sessions colored in gray were omitted.}
  \label{fig:exp123}
\end{figure*}

\section{Related work}

Kenny, Vogt, and others \cite{kenny2007speaker, ivector_good_eer, shriberg2009does, tsuge2018awa} have indeed explored session variability in the context of speaker recognition. Their research is held within the generative methods framework - Gaussian Mixture Models (GMMs), or i-vector \cite{ivector}. In this traditional approach, they tried to represent speaker session variability as a vector or subspace within the GMM model, which was then factorized by the probabilistic LDA \cite{plda} scoring approach. Such explicit modeling of intra speaker variability led to enhanced system performance, for example \cite{ivector_good_2, ivector_good_eer, aronowitz, kenny2008study} show noticeable EER improvement for telephony data. However, it also resulted in a more sophisticated than cosine backend scoring approach \cite{ivector_complexity}, and was subject to testing data domain \cite{kenny2007speaker}. Alternative approaches to model and compensate intra speaker variability within the i-vector framework were done too  \cite{stolcke2007speaker, hayet2014session}, but they didn't present many benefits.

In contrast, the advent of neural network-based speaker recognition systems \cite{xvector, magneto} has shifted how session variability is addressed. In particular, those trained discriminatively with margin losses \cite{am_loss, aam_loss} and used to generate speaker embedding such as x-vector, learn directly from training data. These models are forced to minimize intra speaker variability and maximize inter speaker differences, while implicitly modeling robust speaker clusters at the training stage. As a result, embeddings generated by these networks are robust towards the session variability and allow to simplify scoring backend significantly.

One of the recent studies related to the compensation of speaker session variability \cite{heo2023rethinking} has demonstrated the effectiveness of matching the session embeddings with the help of an additionally trained neural network. Nevertheless, traditional implicit intra speaker variability modelling is the most popular approach nowadays.

\section{Implementation details}

Taking into account the ease of collection and a current trend of pre-training the models on speech in the wild, we were focused on TI speaker recognition.

\subsection{Training}
Based on the problem statement it was decided to use a VoxTube \cite{yakovlev23_interspeech} dataset as a train data. VoxTube stands out for its extensive inter and intra speaker variability, featuring more than 4 million utterances across 10+ languages. We extracted the metadata of each video upload date and used it in our experiments under the assumption of a recording session date. This metadata is available in a Huggingface dataset VoxTube instance \footnote{\url{https://huggingface.co/datasets/voice-is-cool/voxtube}}. In our experiments, we used MUSAN \cite{snyder2015musan} noise corpus for data augmentation and a database of real room impulse responses (RIRs) \cite{Szoke_2019} to apply reverberation to audio signals.

For all experiments we trained ResNet48 architecture in the training setup similar to \cite{yakovlev23_interspeech} with the only difference of reduced batch size twice, down to 256, due to the decreased number of utterances in our experiments.

\subsection{Evaluation}
As a scoring backend, we have used a pairwise cosine scoring with ten 4-second crops, described in \cite{Chung18b}. Models performance evaluation was conducted with the help of Equal Error Rate (EER) metrics. We evaluated the accuracy of models across multiple testing TI datasets covering various languages and environments. Below, we provide a brief description of each dataset:

\begin{itemize}
    \item \textbf{VoxCeleb1} \cite{Nagrani17}:
    The multilingual dataset was collected 'in the wild' from YouTube videos, with a high intra speaker variability achieved by various recording environments, background noises, and other artifacts. We used a Standard-O test protocol, which incorporates 40 speakers.
    \item \textbf{LibriSpeech} \cite{panayotov2015librispeech}:
    The dataset in English, collected from audiobooks, with a low intra speaker variability. We used a \textit{test-clean} subset with 50 speakers and 5.4 hours of speech.
    \item \textbf{FFSVC20} \cite{qin2020ffsvc}: 
    The far-field dataset in the Chinese language, was recorded in the studio, with a high intra speaker variability achieved by various recording devices, distances, and background noises. For evaluation, we used the TI Task 2 development part.
\end{itemize}

For a better readability of a report, we present average EER across all datasets listed above for all of our experiments.

\begin{table}[ht]
    \centering
    \caption{Evaluation metrics for the first and second experiments. Intra spk. - is an average value of ratio of recording sessions dates preserved per speaker relative to the baseline}
    \label{tab:experiment_1_2}
    \scalebox{1}{
    \begin{tabular}{c|c|c|c|c}
        \hline\hline
        \multirow{2}{*}{\textbf{Dataset}} &
        \textbf{Sample} & 
        \multirow{2}{*}{\textbf{Intra spk., \%}} &
        \multicolumn{2}{c}{\textbf{Average}} \\
        &
        \textbf{type} &
        &
        \textbf{EER,\%} &
        \textbf{$\Delta$,\%} \\
        \hline
        
        Baseline                &   -     & 100.0 & 3.40 & -   \\
        \hline
        \multirow{2}{*}{-20\%}  &  tails  & 63.9  & 3.63 & -6.8 \\
                                &  center & 100.0 & 3.45 & -1.5 \\
        \hline
        \multirow{2}{*}{-40\%}  &  tails  & 45.2  & 3.63 & -6.8 \\
                                &  center & 100.0 & 3.59 & -5.6 \\
        \hline
        \multirow{2}{*}{-60\%}  &  tails  & 30.2  & 3.88 & -14.1 \\
                                &  center & 100.0 & 3.36 & 1.2  \\
        \hline\hline
    \end{tabular}
    }
\end{table}

\section{Experiments}

To match the number of utterances across different experiments we randomly sampled half of the utterances for each session in the VoxTube dataset, and used this dataset as a baseline in tables [\ref{tab:experiment_1_2}-\ref{tab:experiment_6}]. This resulted in a baseline containing on average 431 utterances sampled from 60 recording sessions with average intra speaker variability between farthest sessions of 3.6 years for each of 5040 speakers. For the experiments with varying number of speakers, we retained the same language and gender distribution, and hence, can ignore the influence of these factors.

\subsection{Importance of sessions}\label{subsection:importance}

The first question we tried to answer - how important is it to have a wide temporal resolution for speaker recording sessions? For that, we conducted two sets of experiments, where we removed a specific subset of sessions from either the tails or center of the session distribution for each speaker. When removing a specific number of sessions from the dataset we sampled more utterances for the ones preserved in the dataset to obtain a similar number of utterances for all datasets across the experiment. This allows us to compare the models trained with the same training hyperparameters and approximately equal number of utterances used in each experiment.

\subsubsection{Tails removal}\label{subsection:exps_tails}
In this experiment, we focused on dropping sessions that were significantly earlier or later than the median upload date for each speaker, keeping the central temporal range untouched, as illustrated on the plot \textbf{A} of figure \ref{fig:exp123}. We constructed 3 versions of the dataset by excluding a certain percentage of sessions from the tails of the distribution of video upload dates. In the first dataset, we excluded 20\% of sessions (10\% on each side), 40\% in the second dataset, and 60\% in the third dataset.

\subsubsection{Center removal}\label{subsection:exps_center}
In this experiment, we removed the closest sessions for each speaker, as shown in the plot \textbf{B} in the figure \ref{fig:exp123}. We constructed 3 datasets, excluding a certain percentage of sessions from the center of the distribution of video upload dates. Experiments are similar to the ones described in the previous section, with the main difference in excluding the center of the distribution, instead of its tails.

According to the testing results, shown in table \ref{tab:experiment_1_2}, the removal of sessions from the center of the distribution does not affect the model's accuracy significantly, while dropping sessions from the tails of the distribution leads to a significant accuracy degradation. Table \ref{tab:experiment_1_2} column \textbf{Intra spk.} shows the average relative speaker variability preserved in percentages based on video upload dates compared to the baseline. It is calculated as a delta between the furthest videos for each speaker in the dataset. The baseline value of 3.6 years is reduced significantly when dropping distribution tails. Accordingly, it represents the high correlation between intra speaker variability and the model performance.

\begin{table}[h]
    \centering
    \caption{Evaluation metrics for experiments in \ref{subsection:exps_random_and_adaptive_drop}. \#Utt. - is an average number of utterances per speaker, and \#Sess. - is an average number of sessions per speaker.}
    \label{tab:experiment_3_4_5}
    \begin{tabular}{c|c|c|c|c|c}
        \hline\hline
        \multirow{2}{*}{\textbf{Dataset}} &
        \textbf{Sample} & 
        \multirow{2}{*}{\textbf{\# Utt.}} &
        \multirow{2}{*}{\textbf{\# Sess.}} &
        \multicolumn{2}{c}{\textbf{Average}} \\
        &
        \textbf{type} &
        &
        &
        \textbf{EER,\%} &
        \textbf{$\Delta$,\%} \\
        \hline

        Baseline                &     -          & 432                  & 60                  & 3.40 & -    \\
        \hline
        \multirow{3}{*}{-10\%}  &  rand. sess.   & \multirow{2}{*}{390} & \multirow{2}{*}{54} & 3.32 & 2.4   \\
                                &  adapt. sess.  &                      &                     & 3.36 & 1.2   \\
                                \rowcolor{Gray}
                                &  utterance     & 356                  & 57                  & 3.46 & -1.8  \\
        \hline
        \multirow{3}{*}{-20\%}  &  rand. sess.   & \multirow{2}{*}{350} & \multirow{2}{*}{48} & 3.45 & -1.5  \\
                                &  adapt. sess.  &                      &                     & 3.27 & 3.8   \\
                                \rowcolor{Gray}
                                &  utterance     & 319                  & 57                  & 3.33 & 2.1   \\
        \hline
        \multirow{3}{*}{-30\%}  &  rand. sess.   & \multirow{2}{*}{305} & \multirow{2}{*}{42} & 3.46 & -1.8  \\
                                &  adapt. sess.  &                      &                     & 3.34 & 1.8   \\
                                \rowcolor{Gray}
                                &  utterance     & 273                  & 57                  & 3.32 & 2.4   \\
        \hline
        \multirow{3}{*}{-40\%}  &  rand. sess.   & \multirow{2}{*}{265} & \multirow{2}{*}{36} & 3.42 & -0.6  \\
                                &  adapt. sess.  &                      &                     & 3.34 & 1.8   \\
                                \rowcolor{Gray}
                                &  utterance     & 233                  & 57                  & 3.37 & 0.9   \\
        \hline
        \multirow{3}{*}{-50\%}  &  rand. sess.   & \multirow{2}{*}{220} & \multirow{2}{*}{30} & 3.46 & -1.8  \\
                                &  adapt. sess.  &                      &                     & 3.39 & 0.3   \\
                                \rowcolor{Gray}
                                &  utterance     & 200                  & 57                  & 3.32 & 2.4   \\
        \hline
        \multirow{3}{*}{-60\%}  &  rand. sess.   & \multirow{2}{*}{180} & \multirow{2}{*}{24} & 3.73 & -9.7  \\
                                &  adapt. sess.  &                      &                     & 3.39 & 0.3   \\
                                \rowcolor{Gray}
                                &  utterance     & 147                  & 52                  & 3.34 & 1.8   \\
        \hline
        \multirow{3}{*}{-70\%}  &  rand. sess.   & \multirow{2}{*}{135} & \multirow{2}{*}{18} & 3.84 & -12.9 \\
                                &  adapt. sess.  &                      &                     & 3.75 & -10.3 \\
                                \rowcolor{Gray}
                                &  utterance     & 100                  & 46                  & 3.54 & -5.8  \\
        \hline\hline
    \end{tabular}
\end{table}

\begin{figure}[t]
  \hspace*{-0.75cm}
  \centering
  \includegraphics[width=275pt]{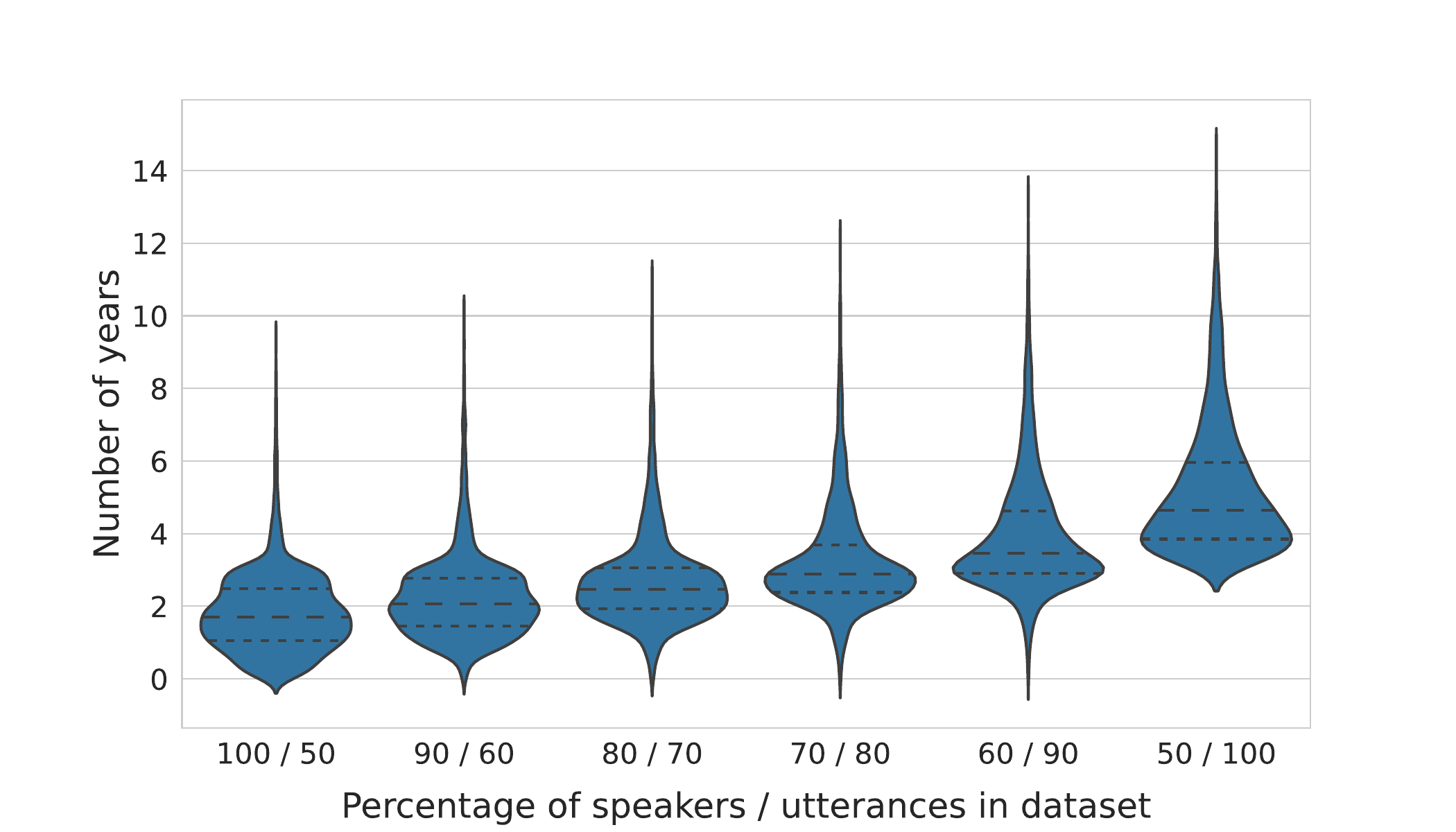}
  \caption{Violin plots of temporal variability per speaker for experiments in \ref{subsection:balance_speakers_sessions}}
  \label{fig:time_delta}
\end{figure}

\subsection{Optimal amount of speech}\label{subsection:exps_random_and_adaptive_drop}

The next question we tried to address - what is the optimal amount of speech and recording sessions per speaker required for model training? For that, we conducted three sets of experiments with various downsampling types.

\subsubsection{Random sessions exclusion}

In this experiment we randomly downsampled the number of sessions for each speaker in the dataset, starting from 10\% up to 70\%, with a 10\% step, as shown on the plot \textbf{C} of figure \ref{fig:exp123}.

\subsubsection{Adaptive sessions exclusion} 

In this experiment we adaptively downsampled the number of sessions for each speaker in the dataset, by removing sessions that stand closest to the neighboring sessions, assuming that they are less informative. We conducted 7 experiments, by reducing the number of sessions from 10\% up to 70\%, with a 10\% step.

\subsubsection{Random utterances exclusion}

In this experiment we randomly reduced the number of utterances in each session by a certain factor, starting from 10\%, and up to 70\%, with a 10\% step. For the big downsampling value of 60\% and more, some of the sessions were discarded completely if no utterances were presented.

According to the testing results shown in table \ref{tab:experiment_3_4_5}, the removal of up to 60\% of the utterances in each session does not affect the model's performance significantly, while starting from the 70\%  downsampling we have noticed a substantial accuracy degradation.

We have also compared different downsampling strategies and noticed that the adaptive session removal works better than the random one due to the better preservation of intra speaker variability. Furthermore, the utterance removal strategy works the best way, due to preserving most of the sessions, therefore, leading to higher intra speaker variability, compared to the session removal approach.

We can conclude that for our sets of experiments with the utterance downsampling strategy, the optimal number of utterances per speaker was at least 147 for 52 recording sessions on average, which is approximately equal to 10 minutes of speech, assuming all utterances are 4 seconds long. We can also note that for the adaptive sessions downsampling strategy significant performance drop appears after the number of utterances and sessions per speaker decreased from 180 and 24 accordingly, which represents 12 minutes of speech per speaker.

\begin{table}[h]
    \centering
    \caption{Evaluation metrics for experiments in \ref{subsection:balance_speakers_sessions}}
    \label{tab:experiment_6}
    \begin{tabular}{c|c|c|c|c}
        \hline\hline
        \multirow{2}{*}{\textbf{Spk., \%}} &
        \multirow{2}{*}{\textbf{Utt., \%}} &
        \textbf{Intra spk.,} &
        \multicolumn{2}{c}{\textbf{Average}}\\
        &
        &
        \textbf{years} &
        \textbf{EER,\%} &
        \textbf{$\Delta$,\%}\\
        \hline
        
        \multicolumn{2}{c|} {Baseline}
                   & 3.6 & 3.40 & -    \\
        \hline
        100  & 50  & 1.9 & 3.67 & -7.9 \\
        \hline
        90   & 60  & 2.4 & 3.46 & -1.8  \\
        \hline
        80  &  70  & 2.8 & 3.53 & -3.8  \\
        \hline
        70  &  80  & 3.4 & 3.61 & -6.2  \\
        \hline
        60  &  90  & 4.1 & 3.77 & -10.9 \\
        \hline
        50  &  100 & 5.4 & 3.76 & -10.6 \\
        \hline\hline
    \end{tabular}
\end{table}

\subsection{Number of speakers vs speakers' temporal variability} \label{subsection:balance_speakers_sessions}

In this set of experiments, we decided to check how the model’s accuracy varies when the number of speakers changes together with their temporal variability. The distribution of time delta between the earliest and latest recording sessions per speaker is shown in figure \ref{fig:time_delta}. We conducted a set of 6 experiments with a fixed number of utterances in each experiment equal to 2.2M, and simultaneously changed the number of speakers and the number of utterances from 100\% of speakers and 50\% of utterances, to 50\% of speakers and 100\% of utterances. We designed our experiments in a way that for experiments with 50\% of speakers, we utilized the ones with the highest temporal variability and removed sessions from distribution tails. For every consecutive experiment, we sampled a portion of utterances uniformly across all the sessions per speaker (see fig. \ref{fig:time_delta}).

According to the testing results shown in table \ref{tab:experiment_6} both, the number of speakers and intra speaker variability is crucial for the model's accuracy. There is a big accuracy drop, when the average intra speaker variability drops below 2 years, even though all speakers were used in the training. When the average intra speaker variability is increased to more than 2 years, it is clear that the number of speakers becomes more important compared to increased intra speaker variability.

\begin{figure}[t]
  \centering
  \includegraphics[width=200pt]{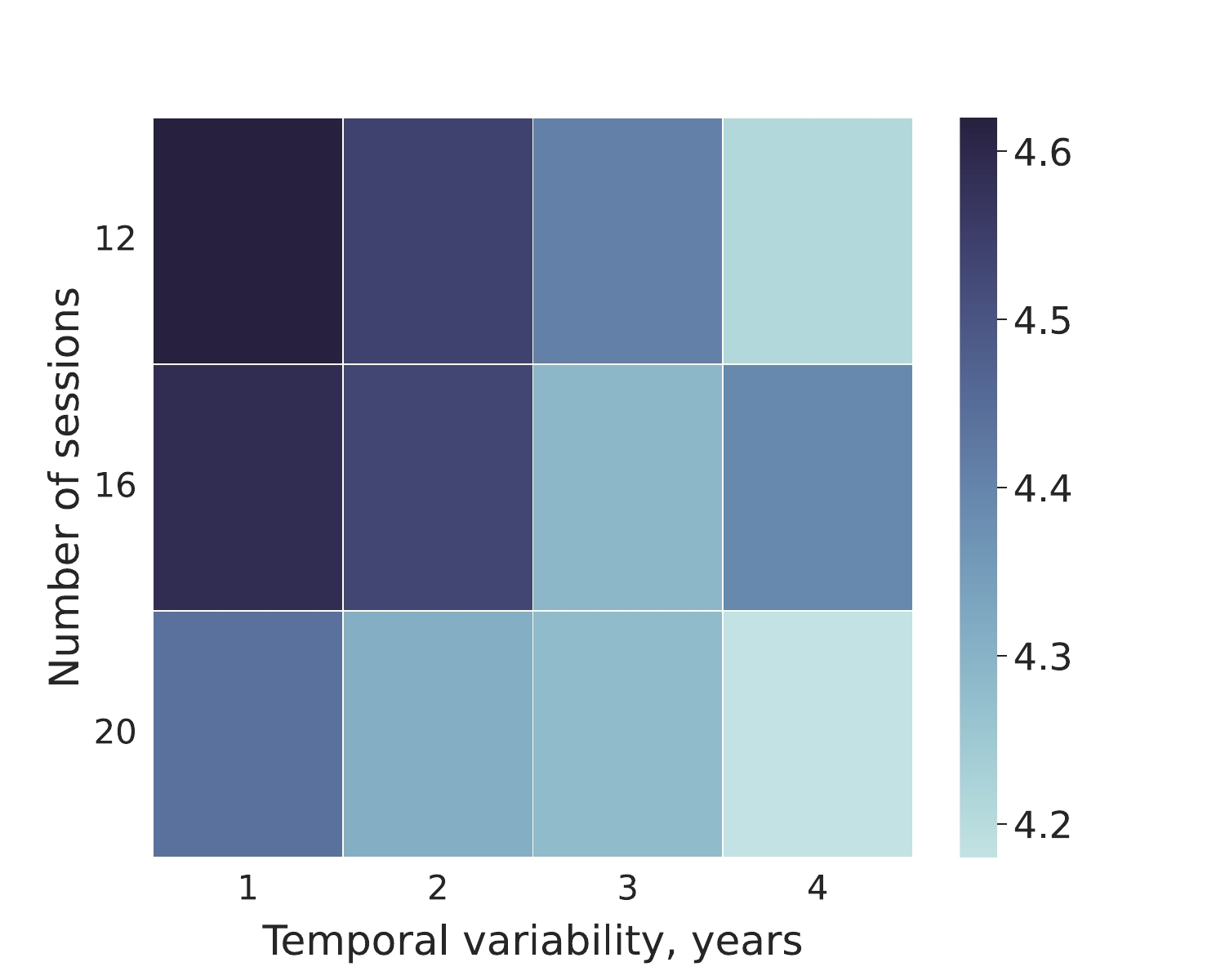}
  \caption{Average EER(\%) for the experiments in \ref{subsection:heatmap_plot}}
  \label{fig:experiment_7}
\end{figure}

\subsection{Temporal variability and the number of sessions} \label{subsection:heatmap_plot}
In this set of experiments, we checked how temporal variability and the number of sessions per speaker trade-off affect the model's accuracy. For that, we selected 4069 speakers with an upload date range from 1 to 4 years, fixed the number of utterances to 120 for each speaker, and conducted 12 experiments, where we changed the number of sessions in the range of 12, 16, and 20 per speaker and the upload range from 1 to 4 years, with a 1-year step.

Testing results for this set of experiments are presented in figure \ref{fig:experiment_7}. According to them, there is a trade-off between the number of sessions and temporal variability of sessions. For instance, one gets similar results for \textit{(20 sessions, 1 year)} and \textit{(12 sessions, 3 years)} datasets, meaning that datasets with 20 sessions per speaker with 1 year temporal variability and 12 sessions per speaker with 3 years temporal variability provide the same results.

\section{Conclusions}

Based on the conducted experiments we can make the following conclusions:

\begin{itemize}
    \item The removal of sessions from the tails of the distribution significantly impacts the model's accuracy, indicating that preserving wide temporal resolution is crucial for maintaining speaker variability. At the same time removing sessions from the center of the distribution has a negligible effect on model accuracy compared to tail removal.
    \item For model training, having at least 24 sessions per speaker (approximately 10 minutes of speech) is optimal. Beyond this point, substantial downsampling of utterances leads to a notable decrease in model accuracy.
    
    \item Both the number of speakers and intra speaker variability significantly affect model accuracy. There is a notable accuracy drop when the average intra speaker variability drops below 2 years, indicating the importance of temporal diversity. However, when intra speaker variability exceeds 2 years, the number of speakers becomes more critical for accuracy.

\end{itemize}

\bibliographystyle{IEEEtran}
\bibliography{voxtube_ablation}

\begin{thebibliography}{10}
\providecommand{\url}[1]{#1}
\csname url@samestyle\endcsname
\providecommand{\newblock}{\relax}
\providecommand{\bibinfo}[2]{#2}
\providecommand{\BIBentrySTDinterwordspacing}{\spaceskip=0pt\relax}
\providecommand{\BIBentryALTinterwordstretchfactor}{4}
\providecommand{\BIBentryALTinterwordspacing}{\spaceskip=\fontdimen2\font plus
\BIBentryALTinterwordstretchfactor\fontdimen3\font minus \fontdimen4\font\relax}
\providecommand{\BIBforeignlanguage}[2]{{%
\expandafter\ifx\csname l@#1\endcsname\relax
\typeout{** WARNING: IEEEtran.bst: No hyphenation pattern has been}%
\typeout{** loaded for the language `#1'. Using the pattern for}%
\typeout{** the default language instead.}%
\else
\language=\csname l@#1\endcsname
\fi
#2}}
\providecommand{\BIBdecl}{\relax}
\BIBdecl

\bibitem{lin2023voxblink}
Y.~Lin, X.~Qin, G.~Zhao, M.~Cheng, N.~Jiang, H.~Wu, and M.~Li, ``Voxblink: A large scale speaker verification dataset on camera,'' \emph{arXiv preprint arXiv:2308.07056}, 2023.

\bibitem{Chung18b}
J.~S. Chung, A.~Nagrani, and A.~Zisserman, ``Voxceleb2: Deep speaker recognition,'' in \emph{INTERSPEECH}, 2018.

\bibitem{zhu2021webface260m}
Z.~Zhu, G.~Huang, J.~Deng, Y.~Ye, J.~Huang, X.~Chen, J.~Zhu, T.~Yang, J.~Lu, D.~Du \emph{et~al.}, ``Webface260m: A benchmark unveiling the power of million-scale deep face recognition,'' in \emph{Proceedings of the IEEE/CVF Conference on Computer Vision and Pattern Recognition}, 2021, pp. 10\,492--10\,502.

\bibitem{HANSEN201894}
\BIBentryALTinterwordspacing
J.~H. Hansen and H.~Bořil, ``On the issues of intra-speaker variability and realism in speech, speaker, and language recognition tasks,'' \emph{Speech Communication}, vol. 101, pp. 94--108, 2018. [Online]. Available: \url{https://www.sciencedirect.com/science/article/pii/S0167639317303849}
\BIBentrySTDinterwordspacing

\bibitem{facerec_intraspeaker}
B.~Moghaddam, ``Principal manifolds and probabilistic subspaces for visual recognition,'' \emph{IEEE Transactions on Pattern Analysis and Machine Intelligence}, vol.~24, no.~6, pp. 780--788, 2002.

\bibitem{voxsrc2023_dku_voxblink}
Z.~Li, Y.~Lin, X.~Qin, N.~Jiang, G.~Zhao, and M.~Li, ``The dku-msxf speaker verification system for the voxceleb speaker recognition challenge 2023,'' 2023.

\bibitem{voxsrc2023_idrnd_voxtube}
N.~Torgashov, R.~Makarov, I.~Yakovlev, P.~Malov, A.~Balykin, and A.~Okhotnikov, ``The id r\&d voxceleb speaker recognition challenge 2023 system description,'' 2023.

\bibitem{yakovlev23_interspeech}
I.~Yakovlev, A.~Okhotnikov, N.~Torgashov, R.~Makarov, Y.~Voevodin, and K.~Simonchik, ``{VoxTube: a multilingual speaker recognition dataset},'' in \emph{Proc. INTERSPEECH 2023}, 2023, pp. 2238--2242.

\bibitem{kenny2007speaker}
P.~Kenny, G.~Boulianne, P.~Ouellet, and P.~Dumouchel, ``Speaker and session variability in gmm-based speaker verification,'' \emph{IEEE Transactions on Audio, Speech, and Language Processing}, vol.~15, no.~4, pp. 1448--1460, 2007.

\bibitem{ivector_good_eer}
R.~Vogt and S.~Sridharan, ``Experiments in session variability modelling for speaker verification,'' in \emph{2006 IEEE International Conference on Acoustics Speech and Signal Processing Proceedings}, vol.~1, 2006, pp. I--I.

\bibitem{shriberg2009does}
E.~Shriberg, S.~Kajarekar, and N.~Scheffer, ``Does session variability compensation in speaker recognition model intrinsic variation under mismatched conditions?'' in \emph{Tenth Annual Conference of the International Speech Communication Association}, 2009.

\bibitem{tsuge2018awa}
S.~Tsuge, S.~Kuroiwa, T.~Ohsuga, and Y.~Ishimoto, ``Awa long-term recorded speech corpus and robust speaker recognition method for session variability,'' in \emph{2018 Oriental COCOSDA-International Conference on Speech Database and Assessments}.\hskip 1em plus 0.5em minus 0.4em\relax IEEE, 2018, pp. 15--20.

\bibitem{ivector}
N.~Dehak, P.~J. Kenny, R.~Dehak, P.~Dumouchel, and P.~Ouellet, ``Front-end factor analysis for speaker verification,'' \emph{IEEE Transactions on Audio, Speech, and Language Processing}, vol.~19, no.~4, pp. 788--798, 2011.

\bibitem{plda}
S.~Ioffe, ``Probabilistic linear discriminant analysis,'' in \emph{Computer Vision -- ECCV 2006}, A.~Leonardis, H.~Bischof, and A.~Pinz, Eds.\hskip 1em plus 0.5em minus 0.4em\relax Berlin, Heidelberg: Springer Berlin Heidelberg, 2006, pp. 531--542.

\bibitem{ivector_good_2}
R.~Vogt, B.~Baker, and S.~Sridharan, ``Modelling session variability in text-independent speaker verification,'' in \emph{Eurospeech/Interspeech: Proceedings of the 9th European Conference on Speech Communication and Technology 2005}.\hskip 1em plus 0.5em minus 0.4em\relax Institut fur Kommunikationsforschung und Ponetik (ISCA), Universitat Bonn, 2005, pp. 3117--3120.

\bibitem{aronowitz}
H.~Aronowitz, D.~Irony, and D.~Burshtein, ``Modeling intra-speaker variability for speaker recognition,'' 09 2005, pp. 2177--2180.

\bibitem{kenny2008study}
P.~Kenny, P.~Ouellet, N.~Dehak, V.~Gupta, and P.~Dumouchel, ``A study of interspeaker variability in speaker verification,'' \emph{IEEE Transactions on Audio, Speech, and Language Processing}, vol.~16, no.~5, pp. 980--988, 2008.

\bibitem{ivector_complexity}
P.~Kenny, G.~Boulianne, P.~Ouellet, and P.~Dumouchel, ``Factor analysis simplified,'' \emph{Acoustics, Speech, and Signal Processing, 1988. ICASSP-88., 1988 International Conference on}, vol.~1, pp. 637--640, 01 2005.

\bibitem{stolcke2007speaker}
A.~Stolcke, S.~S. Kajarekar, L.~Ferrer, and E.~Shrinberg, ``Speaker recognition with session variability normalization based on mllr adaptation transforms,'' \emph{IEEE Transactions on Audio, Speech, and Language Processing}, vol.~15, no.~7, pp. 1987--1998, 2007.

\bibitem{hayet2014session}
D.~Hayet, A.~Radia, D.~Akila, and L.~M. Tayeb, ``Session variability in automatic speaker verification,'' in \emph{2014 International Conference on Multimedia Computing and Systems (ICMCS)}.\hskip 1em plus 0.5em minus 0.4em\relax IEEE, 2014, pp. 185--190.

\bibitem{xvector}
D.~Snyder, D.~Garcia-Romero, G.~Sell, D.~Povey, and S.~Khudanpur, ``X-vectors: Robust dnn embeddings for speaker recognition,'' in \emph{2018 IEEE international conference on acoustics, speech and signal processing (ICASSP)}.\hskip 1em plus 0.5em minus 0.4em\relax IEEE, 2018, pp. 5329--5333.

\bibitem{magneto}
D.~Garcia-Romero, G.~Sell, and A.~Mccree, ``{MagNetO: X-vector Magnitude Estimation Network plus Offset for Improved Speaker Recognition},'' in \emph{Proc. The Speaker and Language Recognition Workshop (Odyssey 2020)}, 2020, pp. 1--8.

\bibitem{am_loss}
F.~Wang, J.~Cheng, W.~Liu, and H.~Liu, ``Additive margin softmax for face verification,'' \emph{IEEE Signal Processing Letters}, vol.~25, no.~7, pp. 926--930, 2018.

\bibitem{aam_loss}
J.~Deng, J.~Guo, N.~Xue, and S.~Zafeiriou, ``Arcface: Additive angular margin loss for deep face recognition,'' in \emph{Proceedings of the IEEE/CVF conference on computer vision and pattern recognition}, 2019, pp. 4690--4699.

\bibitem{heo2023rethinking}
H.-S. Heo, K.~Nam, B.-J. Lee, Y.~Kwon, M.~Lee, Y.~J. Kim, and J.~S. Chung, ``Rethinking session variability: Leveraging session embeddings for session robustness in speaker verification,'' \emph{arXiv preprint arXiv:2309.14741}, 2023.

\bibitem{snyder2015musan}
D.~Snyder, G.~Chen, and D.~Povey, ``Musan: A music, speech, and noise corpus,'' \emph{arXiv preprint arXiv:1510.08484}, 2015.

\bibitem{Szoke_2019}
\BIBentryALTinterwordspacing
I.~Szoke, M.~Skacel, L.~Mosner, J.~Paliesek, and J.~Cernocky, ``Building and evaluation of a real room impulse response dataset,'' \emph{IEEE Journal of Selected Topics in Signal Processing}, vol.~13, no.~4, p. 863–876, Aug 2019. [Online]. Available: \url{http://dx.doi.org/10.1109/JSTSP.2019.2917582}
\BIBentrySTDinterwordspacing

\bibitem{Nagrani17}
A.~Nagrani, J.~S. Chung, and A.~Zisserman, ``Voxceleb: a large-scale speaker identification dataset,'' in \emph{INTERSPEECH}, 2017.

\bibitem{panayotov2015librispeech}
V.~Panayotov, G.~Chen, D.~Povey, and S.~Khudanpur, ``Librispeech: an asr corpus based on public domain audio books,'' in \emph{2015 IEEE international conference on acoustics, speech and signal processing (ICASSP)}.\hskip 1em plus 0.5em minus 0.4em\relax IEEE, 2015, pp. 5206--5210.

\bibitem{qin2020ffsvc}
X.~Qin, M.~Li, H.~Bu, R.~K. Das, W.~Rao, S.~Narayanan, and H.~Li, ``The ffsvc 2020 evaluation plan,'' \emph{arXiv preprint arXiv:2002.00387}, 2020.

\end{thebibliography}

\end{document}